\documentclass[12pt]{article}
\usepackage{cite}
\usepackage{axodraw}
\begin{document}

\title{\bf Renormalization of the energy-momentum tensor in
           noncommutative scalar field theory}

\author{\sc S. Bellucci${}^{a}$,
I.L. Buchbinder${}^{b}$,
V.A. Krykhtin${}^c$\footnote{e-mail: \tt bellucci@lnf.infn.it, joseph@tspu.edu.ru, krykhtin@fnsm.tpu.edu.ru}
\\[0.5cm]
\it ${}^a$INFN, Laboratori Nazionali di Frascati,\\
\it P.O. Box 13, I-00044 Frascati, Italy\\[0.3cm]
\it ${}^b$Department of Theoretical Physics,\\
\it Tomsk State Pedagogical University,\\
\it Tomsk 634041, Russia\\[0.3cm]
\it ${}^c$Laboratory of Mathematical Physics and\\
\it Department of Theoretical and Experimental Physics, \\
\it Tomsk Polytechnic University,\\
\it Tomsk 634050, Russia }

\date{}


\maketitle
\thispagestyle{empty}

\vspace{0.5cm}

\begin{abstract}
We consider the one-loop renormalization of dimension four
composite operators and the energy-momentum tensor
in noncommutative $\phi^4$ scalar field theory.
Proper operator bases are
constructed and it is proved that the bare
composite operators are expressed via renormalized ones, with the
help of a mixing matrix, whose explicit form is calculated.
The corresponding matrix elements turn out to differ from the
commutative theory.
The canonically defined energy-momentum tensor is not finite and
must be replaced by the "improved" one, in order to provide finiteness.
The suitable "improving" terms are found.
\end{abstract}


\section{Introduction}

Noncommutativity of space-time coordinates is an old enough idea
which was proposed by Snyder \cite{Snyder}, in order to
introduce a natural effective ultraviolet cutoff in quantum
field theory.
At present, the interest for field theories with noncommuting
coordinates has grown, owing to the discovery of their relation
to string theory (see \cite{9908142} and references therein),
which is considered as a prototype of the unified theory describing
all fundamental interactions.
Apart from the string theory motivation, noncommutative field
theories seem attractive, as they provide sufficiently consistent non-local
quantum field models. (see the reviews in
\cite{0106048,0111208,0109162}).

Noncommutativity entails some important consequences. One of
them we would like to mention is the so-called UV/IR mixing
\cite{9912072,0002186,0001215,0002075,0201144}. The phenomenon
of mixing UV and IR singularities appears at the quantum level
in nonplanar diagrams: some of the UV singularities of a
commutative theory are converted into IR singularities in its
noncommutative counterpart. So, the contributions of nonplanar
diagrams to the effective action have no UV divergences and the UV
singularities of noncommutative theories are not the same
as those of their commutative counterparts. As a consequence,
this may violate the renormalizability of noncommutative field
theories.

The problem of the renormalizability of noncommutative quantum field
theories was studied in a number of papers for different models
(see e.g.
\cite{9911098,9912075,0005208,0011088,0012024,9912167,0003176,0008057,0207086})
and the general situation concerning the renormalization of
fields and parameters (masses and coupling constants) looks now
clear. Specific features of the renormalization of
supersymmetric models were studied in the papers
\cite{FL,0104190,0002119,0107022,0109222,0102007,0102045,0109130,0110203}.

We would like to point out that, apart from the renormalization
of fields and parameters, there is one more aspect of
renormalization theory, to be considered.
It is the renormalization of composite operators.
As well known, the renormalization of composite operators,
i.e. operator monomials containing products of the fields and their
derivatives at coinciding space-time points, is an independent problem
in quantum field theory.
Even if the Green's functions have been made
finite by the renormalization of fields and parameters, the composite
operators in such a theory are still divergent.
In order to make them finite, one has to carry out an
additional study (see a discussion of this point for commutative
field theory in \cite{Collins}).
As we already mentioned, the renormalization structure of fields and parameters in
noncommutative field theory was studied well enough, but the problem of the
renormalization of composite operators in noncommutative field
theory is still open.
The only paper we know is the work \cite{0004085},
where the renormalization of the operator $\phi^2$ in noncommutative
scalar field theory in four and six dimensions was written out
in an explicit form.

This paper is
devoted to the renormalization of the energy-momentum
tensor in four dimensional noncommutative $\phi^4$ scalar field theory.
The problem under consideration is closely connected with the
renormalization of all dimension four composite operators in the
theory.
Some aspects
concerning the structure of the energy-momentum tensor in noncommutative field
theories were studied in
\cite{0008057,0012112,0103124,0104244,0210288,0212122}, at the
classical level.

General renormalization theory (see e.g. \cite{Collins}) states
that the renormalization of a composite operator demands to take
into account all composite operators of the same mass dimension
and symmetry. This phenomenon is called operator mixing.
We are going to study the operator mixing in noncommutative
field theory and compare it with the situation in the commutative
theory.
The energy-momentum tensor is a linear combination of dimension
4 composite operators and, hence, the corresponding renormalized
tensor is constructed in terms of renormalized composite
operators.
The renormalization of the energy-momentum tensor in the commutative
theory was discussed in details by Brown \cite{Brown}.
An important result asserts that the simultaneous fulfillment of the
finiteness of the energy-momentum tensor and the local conservation law
leads to the necessity to replace the canonical energy-momentum
tensor by the so called "improved" one. We discuss the analogous
situation in the noncommutative theory and show that, unlike the
commutative theory case, the standard conservation law for
noncommutative theories is possible only in the massless case.

The paper is organized as follows.
In the next section we describe the general renormalization procedure
for composite operators.
Section \ref{ComCase} is devoted to a brief review of results
concerning the renormalization of dimension four composite operators in
commutative scalar field theory.
Next, we discuss the renormalization of dimension four composite
operators in the noncommutative theory and emphasize the new
aspects in comparison with the commutative theory.
Section \ref{zeromt} is devoted to the renormalization of
composite operators at zero momentum transfer.
In section \ref{EMT} we consider the
renormalization of the canonical energy-momentum tensor and
prove that the latter must be replaced by the "improved" one,
in order to be made finite.
In section \ref{Summary} we briefly summarize the main results of our work.
The Appendix is devoted to discussing global conservation
laws for the improved energy-momentum tensor.
There, we check the conservation of the energy and momentum which follow
from the finite "improved" energy-momentum tensor and see that the
standard conservation law for noncommutative theories exists
only for the massless theory.

\section{Renormalization scheme for composite operators}

We start this section with a brief formulation of some basic properties of
noncommutative field
theories.
As it is well known, a noncommutative field theory may be
constructed from a
commutative field theory, by replacing the usual product of the fields with the
star one
\begin{eqnarray}
\label{theta}
f\cdot g\to
(f\star g)(x)=
  \left.\exp(\frac{i}{2}\theta^{\mu\nu}\partial_\mu^u\partial_\nu^v)
                f(x+u)g(x+v)\right|_{u=v=0} \neq (g\star f)(x),
\end{eqnarray}
where the constants $\theta^{\mu\nu}$ are noncommutativity
parameters with dimension of a length squared.
For example, for the self-interacting scalar field which we will
study in this paper, the action is written as
\begin{eqnarray}
S=\int d^dx \Bigl[\,
            \frac{1}{2}\partial_\mu\phi\star{}\partial^\mu\phi
            -\frac{m^2}{2}\phi\star{}\phi
            -\frac{\mu^{4-d}\lambda}{4!}\,
              \phi\star{}\phi\star{}\phi\star{}\phi
            \,\Bigr].
\end{eqnarray}
Here $\lambda$ is a dimensionless coupling constant and $\mu$ is
an arbitrary parameter with the dimension of a mass.

The energy-momentum tensor following from this action, with the
help of Noether's theorem, is
\cite{0008057,0012112,0104244}
\begin{eqnarray}
T_{\mu\nu}=\frac{1}{2}\partial_\mu\phi\star{}\partial_\nu\phi
          +\frac{1}{2}\partial_\nu\phi\star{}\partial_\mu\phi
          -\eta_{\mu\nu}
           \Bigl(
            \frac{1}{2}\partial_\alpha\phi\star{}\partial^\alpha\phi
            -\frac{m^2}{2}\phi\star\phi
            -\frac{\mu^{4-d}\lambda}{4!}\phi\star{}\phi\star{}\phi\star{}\phi
           \Bigr).
\end{eqnarray}
In order to construct the renormalized energy-momentum tensor, we need to
renormalize the operators
$\partial_\mu\phi\star{}\partial_\nu\phi$,
$\partial_\alpha\phi\star{}\partial^\alpha\phi$,
$\phi\star{}\phi$, and
$\phi\star{}\phi\star{}\phi\star{}\phi$.

Let us briefly describe the procedure for constructing the
renormalized operator, which is valid for both commutative and
noncommutative theories.
Let $O(\phi)$ be a composite operator, i.e. an operator which is
constructed from products of the fields and their derivatives, at the
same space-time point. In general, the expectation value of such an
operator
\begin{eqnarray}
\label{MV}
<O(\phi)>=\frac{\int D\phi\,e^{iS(\phi)}\,O(\phi)}{\int D\phi\,e^{iS(\phi)}}
\end{eqnarray}
is divergent. For the expectation value (\ref{MV}) to be finite, we need
to renormalize the operator. By definition, a renormalized
operator, denoted as $[O]$, must have a finite expectation value.
The construction of the renormalized operator is performed as follows.
First of all, it is convenient to transform the expression
(\ref{MV}) as follows:
\begin{eqnarray}
\label{MVJ}
<O(\phi)>=
\Biggl(
\frac{\delta}{\delta\,iJ}\,\,
\frac{\int D\phi\,e^{i[S(\phi)+JO(\phi)]}}{\int D\phi\,e^{iS(\phi)}}
\Biggr)\Bigg|_{J=0},
\\
\qquad
JO(\phi)=\int d^dx\, J(x)O(\phi(x)).
\end{eqnarray}
Hereafter, we shall omit the normalizing denominator. Furthermore,
in calculating (\ref{MVJ}) with the
perturbative theory, the UV divergences originate from the
one-particle irreducible (1PI) diagrams.
Hence, if we make all the 1PI diagrams finite, the expectation value of
the composite operator will also be finite.
Therefore, we need to renormalize all the 1PI diagrams of the
theory. The most convenient way of renormalizing the
diagrams is to renormalize the generating functional of the 1PI
diagrams (which is often called the effective action).
For this aim, we apply the following transformations.
Firstly, we introduce the generating functional of the connected
Green's functions $W_J$
\begin{eqnarray}
&&
e^{iW_J(j)}=\int D\phi\,e^{i(S_J(\phi)+j\phi)},
\\
&&\qquad
\label{SJ}
S_J(\phi)=S(\phi)+JO(\phi),
\\
&&\qquad
j\phi=\int d^dx\, j(x)\phi(x).
\end{eqnarray}
Then, one performs the Legendre transform
\begin{eqnarray}
\Gamma_J(\bar\phi)&=&W_J(j(\bar\phi))-j(\bar\phi)\bar\phi,
\end{eqnarray}
where $j(\bar\phi)$ is defined from the equation
\begin{eqnarray}
\frac{{\delta}W_J}{{\delta}j(x)}=\bar\phi(x),
\end{eqnarray}
with $\bar\phi$ being the expectation value of the scalar field $\phi$
in the presence of the sources $j$ and $J$
\begin{eqnarray}
\bar\phi=\frac{\int D\phi\,\phi\, e^{i[S_J(\phi)+j\phi]}}{\int
    D\phi\,e^{i[S_J(\phi)+j\phi]}}.
\end{eqnarray}
As a result of the transformation, we have the equation
\begin{eqnarray}
e^{i\Gamma_J(\bar\phi)}=\int D\phi\,
             e^{i[S_J(\phi)+j(\bar\phi)(\phi-\bar\phi)]}.
\end{eqnarray}
The generating functional of the 1PI Green's functions
$\Gamma_J(\bar\phi)$ is calculated, with the help of
perturbation theory, by means of the loop expansion
\begin{eqnarray}
\Gamma_J(\bar\phi)=\sum_{n=0}^{\infty}\Gamma^{(n)}_J(\bar\phi),
\end{eqnarray}
with $\Gamma^{(n)}_J(\bar\phi)$ being the sum of all the
$n$-loop 1PI diagrams.
Let us shift the integration variable
$\phi\to\varphi=\phi-\bar\phi$. Then we have, in the tree
(zero-loop) approximation,
\begin{eqnarray}
\Gamma^{(0)}_J(\bar\phi)=S_J(\bar\phi).
\end{eqnarray}
The one-loop correction to the $\Gamma^{(0)}_J(\bar\phi)$ reads
\begin{eqnarray}
\label{G1}
\Gamma^{(1)}_J(\bar\phi)=\frac{i}{2}\mbox{\rm tr}\ln
        \frac{\delta^2S_J(\bar\phi)}{\delta\bar\phi(x)\delta\bar\phi(x')}.
\end{eqnarray}
In the following, we will omit the bar sign over $\phi$
($\bar\phi\to\phi$).

The one-loop correction (\ref{G1}) to the effective action is UV
divergent. In order to cancel the divergences, we introduce one-loop
counterterms in the classical action $S_J$ (\ref{SJ})
\begin{eqnarray}
\label{S1}
S_J^{(1)}=S+S_1+JO+JO_1+{\cal O}(J^2),
\end{eqnarray}
where $S_1$ are counterterms which do not contain the source $J$,
and $JO_1$ are counterterms which are linear in $J$.
Note that we need not
calculate counterterms which contain the source $J$ to the second
and higher power, because we need not renormalize expressions like
$O(\phi(x_1))O(\phi(x_2)){\ldots}O(\phi(x_n))$, $n{\geq}2$. As a
result, we may define the renormalized operator as
\begin{eqnarray}
[O]=O+O_1=\sum Z_{0i}O_{0i},
\end{eqnarray}
with $O_{0i}$ being bare (unrenormalized) composite operators and
$Z_{0i}$ being some constants.
Similarly to the commutative case (see e.g. \cite{Collins}), it may
be shown by dimensional analysis that, for renormalizing
a composite operator, in the general case we need all operators of
the same mass dimension.

In this paper we study the renormalization of the
energy-momentum tensor in noncommutative $\phi^4_4$ scalar field
theory. For this purpose, we need to renormalize the dimension four
composite operators which enter into it.
Before considering the noncommutative theory, let us describe the
renormalization of dimension four composite operators in the
corresponding commutative theory.

\section{Basic results on the renormalization of composite
operators in the commutative theory}\label{ComCase}

The renormalization of dimension four composite
operators in the commutative theory was discussed in \cite{Brown}.
As it was shown in that work, composite operators of the same dimension and
Lorentz symmetry are mixed by renormalization.
There are five scalar operators of dimension four, which are
independent of each other.
They are $m_0^4$, $\phi_0^4$, $m_0^2\phi_0^2$,
$\partial^2\phi_0^2$ and
$\partial_\alpha\phi_0\partial^\alpha\phi_0$.
In \cite{Brown} the operator $m_0^4$ was not considered, because
it has no effect on connected Green functions.
Here we ignore this operator, as well.
From the other four operators we can construct an operator,
which is not renormalized.
It is the operator related to the field equation of motion
\begin{eqnarray}
E_0
=
\phi_0\Bigl\{
      \bigl(\partial^2+m_0^2\bigr)\phi_0
      +\frac{1}{3!}\lambda_0\phi_0^3
      \Bigr\}.
\end{eqnarray}
Since it is not renormalized, it is convenient to use it in the
operator basis, instead of some other operator.
Let the operator bases of bare and renormalized
operators be
\begin{eqnarray}
\label{Q}
Q_0=\left(\begin{array}{c}
          \frac{\lambda_0}{4!}\phi_0^4\\[0.1em]
          \frac{1}{2}m_0^2\phi_0^2\\[0.1em]
          E_0\\[0.1em]
          -\partial^2\phi_0^2
          \end{array}
    \right)
& \qquad\mbox{and}\qquad &
[Q]=\left(\begin{array}{c}
          \frac{1}{4!}\mu^{4-d}\lambda [\phi^4]\\[0.1em]
          \frac{1}{2} m^2 [\phi^2 ]\\[0.1em]
          [ E ] \\[0.1em]
          -\partial^2[\phi]
          \end{array}
    \right)
\end{eqnarray}
respectively.
As shown in \cite{Brown}, these bases are related to each other by a
$4\times{}4$ matrix $Z$
\begin{eqnarray}
\label{ZQ}
Q_0=Z[Q],
\end{eqnarray}
having the form
\begin{eqnarray}
\label{Z}
Z=\left(\begin{array}{cccc}
        1+\frac{\lambda^{-1}\beta(\lambda)}{d-4}&\frac{\delta(\lambda)}{d-4}
            &\frac{-\gamma(\lambda)}{d-4}&\frac{f}{d-4}\\
        0&1&0&0\\
        0&0&1&0\\
        0&0&0&z_2
        \end{array}
  \right)
\end{eqnarray}
which is called the mixing matrix.
Here $\beta(\lambda)$, $\gamma(\lambda)$, $\delta(\lambda)$,
$Z_2$ and $f(\lambda)$ are defined from the computation of specific
one-loop diagrams closely connected with the renormalization group equations
\cite{Brown}.
In the noncommutative theory under consideration, we are going to
calculate the analogue of this matrix $Z$ (\ref{Z}).
It is essential for us here that, in calculating the $Z$ matrix
(\ref{Z}), insertions of composite operators in Green's functions
at zero momentum transfer (i.e. operators which are integrated
over space-time) are used.
That is, the method used in \cite{Brown} deals with the renormalization
of the operators $\int{}O(x)d^dx$. On the other hand, in noncommutative field
theories, the operators $\int{}O(x)d^dx$ and $O(x)$ are renormalized
differently. This may be explained in the following way.
Let us denote the Fourier transformed operator as $\tilde{O}(p)$, then
$\int{}O(x)d^dx=\tilde{O}(0)$.
If we insert the operator $\tilde{O}(p)$ in a Green's function and
calculate it using some regularization, then we see that taking the limit
$p\to{}0$ and removing a regulator are non-commuting operations.
Moreover, if we first remove a regulator, then the limit $p\to{}0$
does not exist, due to the contribution of nonplanar diagrams (see
\cite{9912072,0004085}).
By this reason we can repeat the same calculations which are
made in \cite{Brown}, but only for renormalization of composite
operators like $\int{}O(x)d^dx$ and obtain for the dimension-four
composite operators analogue of the $Z$ matrix (\ref{Z}), but with
the renormalization group functions corresponding to the
noncommutative theory (\ref{g}--\ref{b}).
In section \ref{zeromt} we check this conclusion by explicit
calculations.
As far as operators at arbitrary momentum transfer $O(x)$ are
concerned, we cannot use the method of \cite{Brown} for
deriving the structure of the matrix $Z$.
Instead, we are forced to calculate the elements of the $Z$
matrix in an
explicit way, with the help of perturbation theory only,
which we consider in the
one-loop approximation.

\section{One-loop renormalization of dimension four composite
operators in the noncommutative theory}

Next, we investigate in noncommutative $\phi_4^4$ scalar field
theory the renormalization of the dimension four
composite operators which enter into the energy-momentum tensor.
For that, we need  to know the counterterms $S_1$ and $JO_1$ in
(\ref{S1}).
In order to find the counterterms, we use the background field
method, the minimal subtraction scheme and dimensional
regularization. We do not calculate the counterterms
$S_1$ and borrow them from \cite{9912075}. The result of the
renormalization of the scalar field, the mass, and the coupling
constant, in the one-loop approximation, reads
\begin{eqnarray}
\label{g}
\phi_0&=&\phi,
 \qquad\qquad\quad\qquad\qquad\qquad\qquad\gamma(\lambda)=0,
\\
\label{d}
m_0^2
   &=&
      \left(1-\frac{2/3\lambda}{(4\pi)^2(d-4)}\right)\,m^2,
  \qquad\qquad\delta(\lambda)=\frac{2/3\lambda}{(4\pi)^2},
\\
\label{b}
\lambda_0&=&\mu^{4-d}\lambda
      \left(1-\frac{2/3\lambda}{(4\pi)^2(d-4)}\right),
  \qquad\beta(\lambda)=\frac{2/3\lambda^2}{(4\pi)^2},
\end{eqnarray}
with $\gamma(\lambda)$, $\delta(\lambda)$, $\beta(\lambda)$
being the renormalization group functions of the noncommutative
theory.

\begin{figure}[t]
\begin{picture}(455,65)(0,15)
\Vertex(253,30){3.5}
\Vertex(213,30){3.5}
\CArc(233,30)(20,0,360)
\Line(183,29)(213,29)
\Line(183,31)(213,31)
\Line(253,30)(268,50)
\Line(253,30)(268,10)
\Text(173,30)[c]{\mbox{}$J_1$}
\Text(278,50)[c]{\mbox{}$\phi$}
\Text(278,10)[c]{\mbox{}$\phi$}
\end{picture}
\caption{Divergent diagrams linear in $J_1$}\label{2f}
\end{figure}
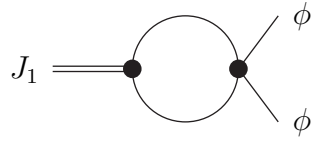

The first operator we renormalize is
$\displaystyle\frac{m^2}{2}\phi\star{}\phi$. The term $JO$ in (\ref{MVJ}) is now
written as
\begin{eqnarray}
\label{1op}
J_1O=\frac{m^2}{2}\int d^dx\, J_1\star{}\phi\star{}\phi.
\end{eqnarray}
In order to simplify the calculations, we perform the Fourier transform of
the field $\phi$ and the source $J$
\begin{eqnarray}
&&
\phi(x)=\int \left(\frac{dp}{2\pi}\right)^d\, e^{ipx}\, \tilde{\phi}(p)
\equiv\int_{p} e^{ipx}\, \tilde{\phi}(p),
\\&&
J(x)=\int \left(\frac{dk}{2\pi}\right)^d\, e^{ikx}\, \tilde{J}(k)
\end{eqnarray}
and work in momentum space.
The divergent diagrams linear in $J_1$ are shown in
Figure~\ref{2f}\footnote{We omit all the counterterms which are
independent of the fields since we assume that the operators are
inserted in the connected Green's functions. The counterterms
which are independent of the fields have no effect on the
connected Green's functions.}.
The diagram in Figure \ref{2f} corresponds to the expression
\begin{eqnarray}
\label{f1b}
&&
\frac{i\mu^{4-d}m^2\lambda}{4!}
 \int_{kp_1p_2}
  \tilde\delta(k+p_1+p_2)\,
  \tilde J_1(k)
  \tilde{\phi}(p_1)
  \tilde{\phi}(p_2)
  e^{-\frac{i}{2}p_1\theta{}p_2}\times
\\
&&\qquad
\times
\int_{p}\frac{2
               +e^{ip\theta{}k}
               +e^{-ip\theta{}k}
               +e^{-ip\theta{}p_1}
               +e^{ip\theta{}p_2}}{(p^2-m^2)((p+k)^2-m^2)}.
\label{mf2}
\end{eqnarray}
As it is known, the expression (\ref{mf2}) may be divided into two parts (see e.g.
\cite{9912072}), which correspond to planar and nonplanar
contributions. The planar contribution
\begin{eqnarray}
\nonumber
&&
\frac{i\mu^{4-d}m^2\lambda}{4!}
 \int_{kp_1p_2}
  \tilde\delta(k+p_1+p_2)\,
  \tilde J_1(k)
  \tilde{\phi}(p_1)
  \tilde{\phi}(p_2)
  e^{-\frac{i}{2}p_1\theta{}p_2}\times
\\
\nonumber
&&\qquad
\times
\int_{p}\frac{2}{(p^2-m^2)((p+k)^2-m^2)}
\\
\nonumber
&&\qquad
=-\frac{2}{(4\pi)^{d/2}}
\frac{i\mu^{4-d}m^2\lambda}{4!}
\Gamma(2-d/2)
 \int_{kp_1p_2}
  \tilde\delta(k+p_1+p_2)\,
  \tilde J_1(k)
  \tilde{\phi}(p_1)
  \tilde{\phi}(p_2)
  e^{-\frac{i}{2}p_1\theta{}p_2}\times
\\
&&\qquad\qquad
\times
\int_0^1dx\, \left[m^2-k^2x(1-x)\right]^{d/2-2}
\end{eqnarray}
has a UV divergence, which is canceled by the counterterm
\begin{eqnarray}
\label{f2ct}
-\frac{1}{d-4}\frac{\lambda}{3!}\frac{m^2}{(4\pi)^2}
\int d^dx\, J_1\star{}\phi\star{}\phi.
\end{eqnarray}
As a result, the renormalized operator in the one-loop
approximation reads

\begin{eqnarray}
\nonumber
\frac{m^2}{2}\left[\phi\star{}\phi\right] &=&
 \left(1-\frac{1/3\lambda}{(4\pi)^2(d-4)}\right)
 \frac{m^2}{2}\phi\star{}\phi
 \\
&=&
 \left(1+\frac{\mu^{d-4}\lambda_0}{3(4\pi)^2(d-4)}\right)
 \frac{m_0^2}{2}
 \phi_0\star\phi_0
 \label{f2}
\end{eqnarray}

\begin{figure}[t]
\begin{picture}(455,65)(0,15)
\CArc(130,30)(20,0,360)
\Vertex(110,30){3.5}
\Line(110,29)(80,29)
\Line(110,31)(80,31)
\Line(110,30)(95,50)
\Line(110,30)(95,10)
\Text(70,30)[c]{\mbox{}$J_2$}
\Text(90,50)[c]{\mbox{}$\phi$}
\Text(90,10)[c]{\mbox{}$\phi$}
\Vertex(355,30){3.5}
\Vertex(315,30){3.5}
\CArc(335,30)(20,0,360)
\Line(285,29)(315,29)
\Line(285,31)(315,31)
\Line(355,30)(370,50)
\Line(315,30)(300,10)
\Line(315,30)(300,50)
\Line(355,30)(370,10)
\Text(275,30)[c]{\mbox{}$J_2$}
\Text(380,50)[c]{\mbox{}$\phi$}
\Text(380,10)[c]{\mbox{}$\phi$}
\Text(295,50)[c]{\mbox{}$\phi$}
\Text(295,10)[c]{\mbox{}$\phi$}
\end{picture}
\caption{Divergent diagrams linear in $J_2$}\label{J2}
\end{figure}
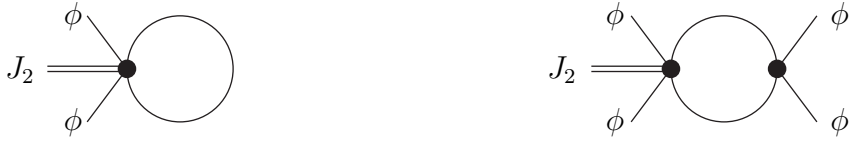

The same scheme is used for renormalizing the other
operators. Let us consider the operator
$\phi\star{}\phi\star{}\phi\star{}\phi$.
For this operator, the term $JO$ in (\ref{MVJ}) is written as
\begin{eqnarray}
J_2O=\mu^{4-d}\int d^dx\,
 J_2\star{}\phi\star{}\phi\star{}\phi\star{}\phi.
\end{eqnarray}
Here $\mu^{4-d}$ is introduced, for the source $J_2$ to be
dimensionless. The divergent diagrams which are linear in $J_2$
are shown in Figure~\ref{J2}.
The expression corresponding to the first diagram in
Figure~\ref{J2} is given by
\begin{eqnarray}
\nonumber
&&
i\mu^{4-d}\int_{kp_1p_2}\,
  \tilde\delta(k+p_1+p_2)\,
  \tilde J_2(k)
  \tilde{\phi}(p_1)
  \tilde{\phi}(p_2)
  e^{-\frac{i}{2}p_1\theta{}p_2}\times
\\
&&\qquad
\times
\int_{p}\frac{3
               +e^{ip\theta{}k}
               +e^{ip\theta{}p_1}
               +e^{ip\theta{}p_2}}{p^2-m^2}.
\label{J21}
\end{eqnarray}
The planar contribution of this diagram reads
\begin{eqnarray}
&&
i\mu^{4-d}\int_{kp_1p_2}\,
  \tilde\delta(k+p_1+p_2)\,
  \tilde J_2(k)
  \tilde{\phi}(p_1)
  \tilde{\phi}(p_2)
  e^{-\frac{i}{2}p_1\theta{}p_2}
  \int_{p}\frac{3}{p^2-m^2}
\\
&&\qquad
 =
 \frac{(\mu/m)^{4-d}}{(4\pi)^{d/2}}3m^2\Gamma(1-d/2)
 \int d^dx \, J_2\star{}\phi\star{}\phi
\end{eqnarray}
and its divergence is canceled by the counterterm
\begin{eqnarray}
\label{ct1f4}
-\frac{1}{d-4}\frac{6m^2}{(4\pi)^2}
 \int d^dx\, J_2\star{}\phi\star{}\phi.
\end{eqnarray}
The remaining part of (\ref{J21}) has no UV divergence.

The divergence of the second diagram in Figure~\ref{J2} is
canceled by the following counterterm:
\begin{eqnarray}
\label{ct2f4}
-\frac{\mu^{4-d}}{d-4}\frac{\lambda}{(4\pi)^2}
 \int d^dx\, J_2\star{}\phi\star{}\phi\star{}\phi\star{}\phi
\end{eqnarray}
and the renormalization of the operator
$\phi\star{}\phi\star{}\phi\star{}\phi$ is given by
\begin{eqnarray}
\nonumber
[\phi\star{}\phi\star{}\phi\star{}\phi]
 &=&
 \left(
 1-\frac{\lambda}{(4\pi)^2(d-4)}
 \right)
 \phi\star{}\phi\star{}\phi\star{}\phi
 -
 \frac{\mu^{d-4}}{d-4}
 \frac{6m^2}{(4\pi)^2}
 \phi\star{}\phi
=
\\
&=&
 \left(
 1-\frac{\mu^{d-4}}{d-4}\frac{\lambda_0}{(4\pi)^2}
 \right)
 \phi_0\star{}
 \phi_0\star{}
 \phi_0\star{}
 \phi_0
 -
 \frac{\mu^{d-4}}{d-4}
 \frac{6m_0^2}{(4\pi)^2}
 \phi_0\star{}
 \phi_0.
\label{f4}
\end{eqnarray}

\begin{figure}[t]
\begin{picture}(455,65)(0,15)
\Vertex(150,30){3.5}
\Vertex(110,30){3.5}
\CArc(130,30)(20,0,360)
\Line(80,29)(110,29)
\Line(80,31)(110,31)
\Line(150,30)(165,50)
\Line(150,30)(165,10)
\Text(70,30)[c]{\mbox{}$J_3$}
\Text(175,50)[c]{\mbox{}$\phi$}
\Text(175,10)[c]{\mbox{}$\phi$}
\Vertex(345,47){3.5}
\Vertex(345,13){3.5}
\Vertex(315,30){3.5}
\CArc(335,30)(20,0,360)
\Line(285,29)(315,29)
\Line(285,31)(315,31)
\Line(345,47)(355,60)
\Line(345,47)(374,50)
\Line(345,13)(355,0)
\Line(345,13)(374,10)
\Text(275,30)[c]{\mbox{}$J_3$}
\Text(363,3)[c]{\mbox{}$\phi$}
\Text(363,60)[c]{\mbox{}$\phi$}
\Text(382,10)[c]{\mbox{}$\phi$}
\Text(382,50)[c]{\mbox{}$\phi$}
\end{picture}
\caption{Divergent diagrams linear in $J_3$}\label{J3}
\end{figure}
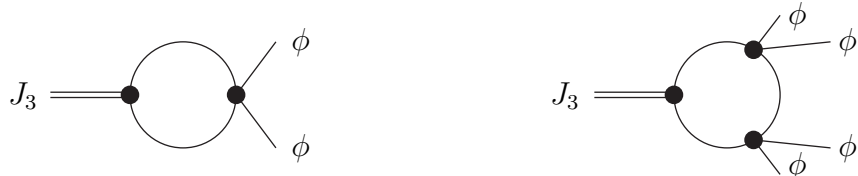

The last operator which we need to renormalize is
$\partial_\mu\phi\star\partial_\nu\phi$. The term $JO$ in
(\ref{MVJ}) corresponding to this operator is written as
\begin{eqnarray}
J_3O=\int d^dx\,
J_3^{\mu\nu}\star\partial_\mu\phi\star\partial_\nu\phi.
\end{eqnarray}
The divergent diagrams which are linear in $J_3$ are shown in
Figure~\ref{J3}.
The counterterms which are needed, in order to cancel their divergences are
\begin{eqnarray}
-\frac{1}{d-4}\,\frac{\lambda}{\,3!\,(4\pi)^2}
 \int d^dx\,
 \left(
 \frac{1}{6}\eta_{\mu\nu}\partial^2J_3^{\mu\nu}
 +
 \frac{1}{3}\partial_\mu\partial_\nu J_3^{\mu\nu}
 +
 m^2\eta_{\mu\nu}J_3^{\mu\nu}
 \right)
 \star{}\phi\star{}\phi
\end{eqnarray}
and
\begin{eqnarray}
-\frac{\mu^{4-d}}{d-4}\,\frac{\lambda^2}{2\,(3!)^2\,(4\pi)^2}
 \int d^dx\,
 \eta_{\mu\nu}J_3^{\mu\nu}\star{}\phi\star{}\phi\star{}\phi\star{}\phi
\end{eqnarray}
respectively. As a result, the renormalized operator is
\begin{eqnarray}
\nonumber
[\partial_\mu\phi\star\partial_\nu\phi]
&=&
\partial_\mu\phi\star\partial_\nu\phi
-
\frac{\mu^{4-d}}{d-4}
\frac{\lambda^2}{2\,(3!)^2\,(4\pi)^2}
\eta_{\mu\nu}\phi\star{}\phi\star{}\phi\star{}\phi
-
\\&&
{}
-
\frac{1}{d-4}\frac{\lambda}{3!(4\pi)^2}
\left(
\frac{1}{6}\eta_{\mu\nu}\partial^2(\phi\star{}\phi)
+
\frac{1}{3}\partial_\mu\partial_\nu(\phi\star{}\phi)
+
m^2\eta_{\mu\nu}\phi\star{}\phi
\right)
\\
\nonumber
&=&
\partial_\mu\phi_0\star\partial_\nu\phi_0
-
\frac{\mu^{d-4}}{d-4}
\frac{\lambda_0^2}{2\,(3!)^2\,(4\pi)^2}
\eta_{\mu\nu}\phi_0\star{}\phi_0\star{}\phi_0\star{}\phi_0
-
\\&&{}
\hspace*{-2em}
-
\frac{\mu^{d-4}}{d-4}\frac{\lambda_0}{3!(4\pi)^2}
\left(
\frac{1}{6}\eta_{\mu\nu}\partial^2(\phi_0\star{}\phi_0)
+
\frac{1}{3}\partial_\mu\partial_\nu(\phi_0\star{}\phi_0)
+
m_0^2\eta_{\mu\nu}\phi_0\star{}\phi_0
\right).
\label{d2}
\end{eqnarray}
As a consequence, we have
\begin{eqnarray}
[\partial_\alpha\phi\star\partial^\alpha\phi]
&=&
\nonumber
\partial_\alpha\phi_0\star\partial^\alpha\phi_0
-
\frac{\mu^{d-4}}{d-4}
\frac{2\lambda_0^2}{(3!)^2\,(4\pi)^2}
\phi_0\star{}\phi_0\star{}\phi_0\star{}\phi_0
-
\\&&{}
-
\frac{\mu^{d-4}}{d-4}\frac{\lambda_0}{3!(4\pi)^2}
\left(
\partial^2(\phi_0\star{}\phi_0)
+
4m_0^2\phi_0\star{}\phi_0
\right).
\label{d3}
\end{eqnarray}
For convenience in the following calculations,
we convert relations (\ref{f2}), (\ref{f4}), (\ref{d2}) and
(\ref{d3}) and express the bare operators in terms of the renormalized
ones
\begin{eqnarray}
\label{bf2}
\frac{m_0^2}{2}
 \phi_0\star{}\phi_0
&=&
\left(1-\frac{1/3\lambda}{(4\pi)^2(d-4)}\right)
   \frac{m^2}{2}\left[\phi\star{}\phi\right],
\\
\nonumber
\frac{\lambda_0}{4!}
 \phi_0\star{}
 \phi_0\star{}
 \phi_0\star{}
 \phi_0
 &=&
\frac{\mu^{4-d}\lambda}{4!}
 \left(
 1+\frac{1/3\lambda}{(4\pi)^2(d-4)}
 \right)
[\phi\star{}\phi\star{}\phi\star{}\phi]
\\&&\qquad{}
+
 \frac{1}{d-4}
 \frac{\lambda}{2\,(4\pi)^2}
 \frac{m^2}{2}[\phi\star{}\phi],
\label{bf4}
\\
\nonumber
\partial_\mu\phi_0\star\partial_\nu\phi_0
&=&
[\partial_\mu\phi\star\partial_\nu\phi]
+
\frac{\mu^{4-d}}{d-4}
\frac{\lambda^2}{2\,(3!)^2\,(4\pi)^2}
\eta_{\mu\nu}[\phi\star{}\phi\star{}\phi\star{}\phi]
+
\\&&{}
\hspace*{-3em}
+
\frac{1}{d-4}\frac{\lambda}{3!(4\pi)^2}
\left(
\frac{1}{6}\eta_{\mu\nu}\partial^2[\phi\star{}\phi]
+
\frac{1}{3}\partial_\mu\partial_\nu[\phi\star{}\phi]
+
m^2\eta_{\mu\nu}[\phi\star{}\phi]
\right),
\label{pfpf}
\\
\nonumber
\partial_\alpha\phi_0\star\partial^\alpha\phi_0
&=&
[\partial_\alpha\phi\star\partial^\alpha\phi]
+
\frac{\mu^{4-d}}{d-4}
\frac{2\lambda^2}{(3!)^2\,(4\pi)^2}
[\phi\star{}\phi\star{}\phi\star{}\phi]
+
\\&&{}
+
\frac{1}{d-4}\frac{\lambda}{3!(4\pi)^2}
\left(
\partial^2[\phi\star{}\phi]
+
4m^2[\phi\star{}\phi]
\right).
\label{pf2}
\end{eqnarray}
For comparison with the commutative case, it is convenient to
write down the renormalization of dimension four scalar composite
operators like (\ref{ZQ}). First of all, it should be noted that
there are more independent scalar operators in the
noncommutative theory, because the fields do not commute and, for
example, the operators $\phi_0\star\partial^2\phi_0$ and
$(\partial^2\phi_0)\star\phi_0$ are independent. We choose the
bases as
\begin{eqnarray}
\label{QNC}
Q^{nc}_0=\left(\begin{array}{c}
          \frac{\lambda_0}{4!}\phi_0\star\phi_0\star\phi_0\star\phi_0\\[0.1em]
          \frac{1}{2}m_0^2\phi_0\star\phi_0\\[0.1em]
          \phi_0\star{}L^{nc}_0\\[0.1em]
          L^{nc}_0\star\phi_0\\[0.1em]
          \partial^2(\phi_0\star\phi_0)
          \end{array}
    \right)
& \qquad\mbox{and}\qquad &
[Q^{nc}]=\left(\begin{array}{c}
          \frac{1}{4!}\mu^{4-d}\lambda
               [\phi\star\phi\star\phi\star\phi]\\[0.1em]
          \frac{1}{2} m^2 [\phi\star\phi]\\[0.1em]
          [ \phi\star{}L^{nc} ] \\[0.1em]
          [ L^{nc}\star\phi ] \\[0.1em]
          \partial^2[\phi\star\phi]          \end{array}
    \right).
\end{eqnarray}
Here $L_0^{nc}$ and $L^{nc}$ are the field equations
\begin{eqnarray}
\label{L0}
L_0^{nc}
&=&
(\partial^2+m_0^2)\phi_0+\frac{\lambda_0}{3!}\phi_0\star\phi_0\star\phi_0,
\\
L^{nc}
&=&
(\partial^2+m^2)\phi+\frac{\mu^{4-d}\lambda}{3!}\phi\star\phi\star\phi.
\end{eqnarray}
The bases $Q_0^{nc}$ and $[Q^{nc}]$ are related between themselves by the equation
\begin{eqnarray}
\label{ZQNC}
Q^{nc}_0=Z^{nc}[Q^{nc}],
\end{eqnarray}
where the matrix $Z^{nc}$ looks as follows
\begin{eqnarray}
\label{NCZ}
Z^{nc}=\left(\begin{array}{ccccc}
        1+\frac{1/3\lambda}{(4\pi)^2(d-4)}
          &\frac{\lambda}{2\,(4\pi)^2(d-4)}
            &0&0&0\\
        0&1-\frac{1/3\lambda}{(4\pi)^2(d-4)}&0&0&0\\
              0&0&1&0&0\\
              0&0&0&1&0\\
              \\
        0&0&0&0&1+\frac{1/3\lambda}{(4\pi)^2(d-4)}
        \end{array}
  \right).
\end{eqnarray}
Here, we see that the elements of the mixing matrix in the
noncommutative theory (\ref{NCZ})
are not related with renormalization group functions in the same way,
as in the commutative case (\ref{Z}), what confirms our assumption
made at the end of section \ref{ComCase}.
Now, we have all operators needed for the construction of the
energy-momentum tensor.

\section{Renormalization of dimension four composite operators at
zero momentum transfer}\label{zeromt}

We have shown in the previous section that the renormalization of
dimension four operators in the noncommutative theory differs very
much from that in the commutative theory.
The operator basis (\ref{QNC}) consists of five operators unlike the
commutative case (\ref{Q}).
In order to find any similarities between the commutative and
noncommutative cases, we consider the renormalization of dimension
four operators at zero momentum transfer.
Such operators are defined as space-time integrals of local composite
operators.
In this case the number of independent operators is two less
than in (\ref{QNC}).
First of all, because of the cyclicity property
\begin{eqnarray}
\label{cp}
\int{}(f\star{}g)(x)\,d^dx
&=&
\int{}(g\star{}f)(x)\,d^dx,
\end{eqnarray}
the third and the fourth operators in (\ref{QNC}), being
integrated over space-time coincide.
Secondly, the last operator in (\ref{QNC}) vanishes when it is
integrated over space-time.
So, we have only three operators in the bases
\begin{eqnarray}
\label{QNC0}
Q'_0=\left(\begin{array}{c}
                \frac{\lambda_0}{4!}
                \int\phi_0\star\phi_0\star\phi_0\star\phi_0\,d^dx\\[0.1em]
          \frac{m^2_0}{2}\int\phi_0\star\phi_0\,d^dx\\[0.1em]
          \int\phi_0\star{}L^{nc}_0\,d^dx
          \end{array}
    \right)
& \hspace{0.1em}\mbox{and}\hspace{0.1em} &
[Q']=\left(\begin{array}{c}
          \frac{\mu^{4-d}\lambda}{4!}
               [\int\phi\star\phi\star\phi\star\phi\,d^dx]\\[0.1em]
          \frac{m^2}{2} [\int\phi\star\phi\,d^dx]\\[0.1em]
          [\int \phi\star{}L^{nc}\,d^dx ]
          \end{array}
    \right).
\end{eqnarray}
It is easy to see that the same number of basis operators will
also occur in the commutative case after integrating (\ref{Q}).
Next, we follow the procedure formulated in \cite{Brown},
adapting it for a noncommutative field theory.
We begin with the general relations
\begin{eqnarray}
\mu\frac{d\lambda}{d\mu}
&=&
(d-4)\lambda+\beta(\lambda),
\\
\mu\frac{dm^2}{d\mu}
&=&
m^2\delta(\lambda),
\end{eqnarray}
where $\beta(\lambda)$, $\delta(\lambda)$ are the
renormalization group functions in noncommutative theory.

Let us consider the renormalization of the operator
$m_0^2\int\phi_0\star\phi_0d^dx$.
Since
$\int\phi_0\star\phi_0d^dx$ is a dimension two operator, then by
dimensional analysis it can be mixed by renormalization with
dimension two operators.
But this operator is the only dimension two operator which
depends on the fields and preserves $\phi\to-\phi$ symmetry.
So, we may expect that
\begin{eqnarray}
\label{Z2}
m_0^2\int\phi_0\star\phi_0\,d^dx
&=&
z(\lambda)
\Bigl[
m^2\int\phi\star\phi{}\,d^dx
\Bigr],
\end{eqnarray}
where $z$ is some renormalization constant, which has
the structure $z=1+\mbox{\it{}divergent}\,\,\mbox{\it{}terms}$.

Let us consider the renormalized Green's function
\begin{eqnarray}
\label{G}
G_N(x_1,\ldots,x_N)
&=&
z_\phi^{-N}
\int{}D\phi_0\,\phi_0(x_1)\ldots{}\phi_0(x_N)\,e^{iS},
\end{eqnarray}
where $z_\phi$ is the field renormalization constant
$\phi_0=z_{\phi}\phi$.
Its derivative with respect to the renormalized parameters is a
finite quantity.
Let us consider its derivative with respect to the renormalized
mass\footnote{We omit the normalizing denominator in (\ref{G})
and subsequent expressions. The derivatives acting on
the denominator produce disconnected graphs which we ignore.}
\begin{eqnarray}
\nonumber
m^2\frac{\partial}{\partial{}m^2}
G_N(x_1,\ldots,x_N)
&=&
z_\phi^{-N}
\int{}D\phi_0\,\phi_0(x_1)\ldots{}\phi_0(x_N)\,
m_0^2\frac{\partial{}iS}{\partial{}m_0^2}\,e^{iS}
\\
\nonumber
&=&
z_\phi^{-N}
\int{}D\phi_0\,\phi_0(x_1)\ldots{}\phi_0(x_N)
\left(\int\frac{-im_0^2}{2}\phi_0\star\phi_0\,d^dx\right)e^{iS}
\\
&=&
G_N(x_1,\ldots,x_N;\int\frac{-im_0^2}{2}\phi_0\star\phi_0\,d^dx).
\label{G2}
\end{eqnarray}
The l.h.s. of this expression is finite, hence the r.h.s. of this
expression must also be finite.
Substituting, instead of the bare operator, its expression from
the renormalized one (\ref{Z2}) in the last line of (\ref{G2})
and recalling that this expression is finite, we find that $z_2$ has
no divergent terms. Thus, we have
\begin{eqnarray}
m_0^2\int\phi_0\star\phi_0\,d^dx
&=&
\Bigl[
m^2\int\phi\star\phi{}\,d^dx
\Bigr].
\label{zerof^2}
\end{eqnarray}

Let us consider the operator
\begin{math}
E_0
=
\int\phi_0\star{}L_0^{nc}\,d^dx
=
-\int\phi_0(x)\cdot
\frac{\displaystyle\delta{}S}{\displaystyle\delta\phi_0(x)}\,d^dx.
\end{math}
Inserting it into the renormalized Green's function (\ref{G}) yields
\begin{eqnarray}
\nonumber
G_N(x_1,\ldots,x_N;E_0)
&=&
z_\phi^{-N}
\int{}D\phi_0\,\phi_0(x_1)\ldots{}\phi_0(x_N)
\left(
i\int\phi_0(x)
\frac{\delta{}iS}{\delta\phi_0(x)}\,d^dx\right)
\,e^{iS}
\\
&=&
z_\phi^{-N}
\int{}d^dx
\int{}D\phi_0\,\phi_0(x_1)\ldots{}\phi_0(x_N)
i\phi_0(x)
\frac{\delta{}e^{iS}}{\delta\phi_0(x)}.
\end{eqnarray}
Integrating the functional integral by parts and neglecting the
term proportional to $\delta(0)$, which is supposed to be zero in
dimensional regularization, we get
\begin{eqnarray}
\label{GN}
G_N(x_1,\ldots,x_N;E_0)
&=&
-iN
G_N(x_1,\ldots,x_N).
\end{eqnarray}
Thus, the operator $E_0$ is finite
\begin{eqnarray}
\int\phi_0\star{}L_0^{nc}\,d^dx
&=&
\Bigl[
\int\phi\star{}L^{nc}\,d^dx
\Bigr].
\end{eqnarray}

Let us consider the derivative of the renormalized Green's function
(\ref{G}) with respect to $\lambda$
\begin{eqnarray}
\nonumber
\frac{\partial{}G}{\partial\lambda}
&=&
\frac{-i}{(d-4)\lambda+\beta(\lambda)}
\Biggl\{
iN\gamma(\lambda)\,G(x_1,\ldots,x_N)
\\
\nonumber
&&\qquad{}
+
(d-4)\,
G(x_1,\ldots,x_N;
  \frac{\lambda_0}{4!}\int\phi_0\star\phi_0\star_0\star\phi_0\,d^dx)
\\
&&\qquad{}
-
\delta(\lambda)\,
G(x_1,\ldots,x_N;
  \frac{m^2_0}{2}\int\phi_0\star\phi_0\,d^dx)
\Biggr\}.
\label{L}
\end{eqnarray}
Here $\gamma(\lambda)$ is the renormalization group function which
is defined by
\begin{eqnarray}
\mu\frac{dz_\phi}{d\mu}
&=&
\gamma(\lambda)\,z_{\phi}.
\end{eqnarray}
Using (\ref{zerof^2}, \ref{GN}) and substituting in (\ref{L})
\begin{eqnarray}
\nonumber
\frac{\lambda_0}{4!}
\int\phi_0\star\phi_0\star\phi_0\star\phi_0\,d^dx
&=&
a\frac{\mu^{4-d}\lambda}{4!}
\Bigl[
\int\phi\star\phi\star\phi\star\phi\,d^dx
\Bigr]
\\
&&{}
+
b\frac{m^2}{2}
\Bigl[\int\phi\star\phi\,d^dx\Bigr]
+
c\Bigl[\int\phi\star{}L^{nc}\,d^dx\Bigr],
\end{eqnarray}
where $a$, $b$, $c$ are some renormalization constants to be
defined, we find that the following quantities must be finite:
\begin{eqnarray}
\frac{(d-4)a}{(d-4)\lambda+\beta(\lambda)},
\qquad
\frac{(d-4)b-\delta(\lambda)}{(d-4)\lambda+\beta(\lambda)},
\qquad
\frac{(d-4)c+\gamma(\lambda)}{(d-4)\lambda+\beta(\lambda)}.
\end{eqnarray}
Expanding the denominator in a power series in $\beta(\lambda)$,
producing an ascending series in $(d-4)$, we find
\begin{eqnarray}
a=1+\frac{\lambda^{-1}\beta(\lambda)}{d-4},
\qquad
b=\frac{\delta(\lambda)}{d-4},
\qquad
c=\frac{-\gamma(\lambda)}{d-4}.
\end{eqnarray}

Ultimately, we have
\begin{eqnarray}
Q'_0=Z'[Q'],
\end{eqnarray}
where $Z'$ is the following matrix:
\begin{eqnarray}
\label{Z'}
Z'=\left(\begin{array}{ccc}
        1+\frac{\lambda^{-1}\beta(\lambda)}{d-4}
          &\frac{\delta(\lambda)}{d-4}
            &\frac{-\gamma(\lambda)}{d-4}\\
        0&1&0\\
        0&0&1\\
        \end{array}
  \right).
\end{eqnarray}
Here $\beta(\lambda)$, $\gamma(\lambda)$, $\delta(\lambda)$
are the renormalization group functions of the noncommutative
theory, whose explicit form in the one-loop approximation is
given by (\ref{g}--\ref{b}).
Next, we confirm this statement by explicit one-loop calculations.

Let us consider the renormalization of the operator
$\frac{m^2}{2}\int\phi\star\phi\,d^dx$.
This situation corresponds to the case $J_1(x)=1$ in (\ref{1op})
or in momentum space $\tilde{J}_1(k)=(2\pi)^d\delta(k)$.
Substituting this expression for the source in (\ref{f1b}) we
find that the number of planar diagrams is twice that for
the operator at arbitrary momentum transfer. So, the corresponding
counterterm is also two times bigger than (\ref{f2ct}).
As a result, we obtain
\begin{eqnarray}
\nonumber
\frac{m^2}{2}\left[\int\phi\star{}\phi\,d^dx\right] &=&
 \left(1-\frac{2/3\lambda}{(4\pi)^2(d-4)}\right)
 \frac{m^2}{2}\int\phi\star{}\phi\,d^dx
\\
\label{mff}
&=&
 \frac{m_0^2}{2}\int\phi_0\star\phi_0\,d^dx.
\end{eqnarray}
This result just corresponds to the second line of the matrix
(\ref{Z'}).

In order to check the first line of (\ref{Z'}), we consider
the renormalization of the operator
$\int\phi\star{}\phi\star{}\phi\star{}\phi\,d^dx$.
Carrying out an analogous procedure as for the previous operator, we get that
both counterterms which are needed to cancel the UV divergences
of the operator at zero momentum transfer are a factor 4/3 larger than the
corresponding counterterms at arbitrary momentum transfer
(\ref{ct1f4}) and (\ref{ct2f4}).
Hence, we have
\begin{eqnarray}
\nonumber
\left[\int\phi\star{}\phi\star{}\phi\star{}\phi\,d^dx\right]
 &=&
 \left(
 1-\frac{4/3\lambda}{(4\pi)^2(d-4)}
 \right)
 \int\phi\star{}\phi\star{}\phi\star{}\phi\,d^dx
\\
&&{}
 -
 \frac{\mu^{d-4}}{d-4}
 \frac{8m^2}{(4\pi)^2}
 \int\phi\star{}\phi\,d^dx
\end{eqnarray}
and then
\begin{eqnarray}
\nonumber
\frac{\lambda_0}{4!}
\int\phi_0\star\phi_0\star\phi_0\star\phi_0\,d^dx
&=&
\left(1+\frac{1}{d-4}\frac{2/3\lambda}{(4\pi)^2}\right)
  \frac{\mu^{4-d}\lambda}{4!}
\left[
  \int\phi\star\phi\star\phi\star\phi\,d^dx\right]
\\
\label{ffff}
&&{}
+
\frac{1}{d-4}\frac{2/3\lambda}{(4\pi)^2}
\frac{m^2}{2}\left[\int\phi\star\phi\,d^dx\right].
\end{eqnarray}
Taking into account the renormalization group functions of the
noncommutative theory (\ref{g}--\ref{b}) this renormalization
relation corresponds to the first line of the matrix (\ref{Z'}).

Finally, it is a simple exercise to prove that
\begin{eqnarray}
\nonumber
\int\phi_0\star\partial^2\phi_0\,d^dx
&=&
-\int\partial_\alpha\phi_0\star\partial^\alpha\phi_0\,d^dx
\\
&=&
\label{fppf}
\left[\int\phi\star\partial^2\phi\,d^dx\right]
-
\frac{1}{d-4}\,\frac{8/3\lambda}{(4\pi)^2}\,
\frac{m^2}{2}\left[\int\phi\star\phi\,d^dx\right]
\\
\nonumber
&&{}
-
\frac{1}{d-4}\,\frac{8/3\lambda}{(4\pi)^2}\,
\frac{\mu^{4-d}\lambda}{4!}\,
\left[\int\phi\star\phi\star\phi\star\phi\,d^dx\right].
\end{eqnarray}
Collecting together (\ref{mff}), (\ref{ffff}), (\ref{fppf}),
we find that
\begin{eqnarray}
\int\phi_0\star{}L_0^{nc}\,d^dx
=
\left[\int\phi\star{}L^{nc}\,d^dx\right],
\end{eqnarray}
which completes our check that the mixing matrix for
dimension four composite operators at arbitrary momentum
transfer is of the form given in (\ref{Z'}).
Thus,we see that the renormalization of the composite operators
at zero momentum transfer in noncommutative and commutative
theories are very similar in contrast to the local composite operators.

\section{One-loop renormalization of the energy-momentum tensor}\label{EMT}

In this section we study the problem of constructing the finite
operator corresponding to the energy-momentum tensor.
First of all, we want to check whether the bare
energy-momentum tensor in noncommutative $\phi^4_4$ scalar field
theory is finite or not.
In terms of bare fields, the latter is given
by
\begin{eqnarray}
\nonumber
T_{0\mu\nu}
&=&
\frac{1}{2}\partial_\mu\phi_0\star{}
           \partial_\nu\phi_0
+\frac{1}{2}\partial_\nu\phi_0\star{}
            \partial_\mu\phi_0
\\&&{}
-\eta_{\mu\nu}
   \Bigl(
       \frac{1}{2}\partial_\alpha\phi_0\star{}
                  \partial^\alpha\phi_0
       -\frac{m_0^2}{2}\,\phi_0\star\phi_0
       -\frac{\lambda_0}{4!}\,\phi_0\star{}
                                   \phi_0\star{}
                                   \phi_0\star{}
                                   \phi_0
    \Bigr).
\label{bemt}
\end{eqnarray}
In the previous section we found the renormalization  of all
the composite operators (\ref{bf2}--\ref{pf2}), which enter into the bare
energy-momentum tensor (\ref{bemt}).
Substituting (\ref{bf2}--\ref{pf2}) into eq. (\ref{bemt}), we see
that the operator $T_{0\mu\nu}$ is not finite
\begin{eqnarray}
\nonumber
T_{0\mu\nu}
&=&
[T_{\mu\nu}]
-
\frac{1}{d-4}\,\frac{\lambda}{3!\,(4\pi)^2}\,\eta_{\mu\nu}\,
  \frac{m^2}{2}[\phi\star{}\phi]
\\&&\qquad{}
+
\frac{1}{d-4}\,\frac{\lambda}{18(4\pi)^2}\,
(\partial_\mu\partial_\nu-\eta_{\mu\nu}\partial^2)
[\phi\star{}\phi].
\label{tdiv}
\end{eqnarray}
Here
\begin{eqnarray}
\nonumber
[T_{\mu\nu}]
&=&
\frac{1}{2}[\partial_\mu\phi\star{}
           \partial_\nu\phi]
+\frac{1}{2}[\partial_\nu\phi\star{}
            \partial_\mu\phi]
\\&&\qquad
-\eta_{\mu\nu}
   \Bigl(
       \frac{1}{2}[\partial_\alpha\phi\star{}
                  \partial^\alpha\phi]
       -\frac{m^2}{2}[\phi\star\phi]
       -\frac{\mu^{4-d}\lambda}{4!}[\phi\star{}
                                   \phi\star{}
                                   \phi\star{}
                                   \phi]
    \Bigr).
\end{eqnarray}
As in the case of the commutative scalar field theory,
$T_{0\mu\nu}$ needs
to be improved, in order to be made finite \cite{Brown}.
For this aim, we add to $T_{0\mu\nu}$ (\ref{bemt}) all possible
independent dimension-four composite operators, which are symmetric
tensors of rank two, with arbitrary coefficients.
The operators are all the scalar ones in $Q_0$
(\ref{QNC}) times $\eta_{\mu\nu}$, with the addition of
three additional tensor operators, i.e.
\begin{math}
\partial_\mu\phi_0\star\partial_\nu\phi_0
+\partial_\mu\phi_0\star\partial_\nu\phi_0,
\end{math}
\begin{math}
\partial^2_{\mu\nu}\phi_0\star\phi_0,
\end{math}
\begin{math}
\phi_0\star\partial^2_{\mu\nu}\phi_0.
\end{math}
So, we have the expression
\begin{eqnarray}
\nonumber
T'_{0\mu\nu}
&=&
T_{0\mu\nu}
+
\frac{A}{2}\left(
   \partial_\mu\phi_0\star\partial_\nu\phi_0
  +\partial_\mu\phi_0\star\partial_\nu\phi_0
\right)
+B\eta_{\mu\nu}L_0^{nc}\star\phi_0
+C\eta_{\mu\nu}\phi_0\star{}L_0^{nc}
\\
&&\qquad{}
+D(\partial^2_{\mu\nu}\phi_0)\star\phi_0
+E\phi_0\star\partial^2_{\mu\nu}\phi_0
+F\eta_{\mu\nu}\frac{\lambda_0}{4!}
  \phi_0\star\phi_0\star\phi_0\star\phi_0
\\
\nonumber
&&\qquad{}
+G\eta_{\mu\nu}\frac{m_0^2}{2}\phi_0\star\phi_0
+H\eta_{\mu\nu}\partial^2(\phi_0\star\phi_0),
\end{eqnarray}
which we want to make finite.
Using eqs. (\ref{bf2}--\ref{pf2}), (\ref{ZQNC}), (\ref{tdiv})
and the following renormalization relations:
\begin{eqnarray}
\nonumber
(\partial^2_{\mu\nu}\phi_0)\star\phi_0
&=&
[(\partial^2_{\mu\nu}\phi)\star\phi]
-
\frac{\mu^{4-d}}{d-4}\frac{\lambda^2}{2\,3!\,(4\pi)^2}
\eta_{\mu\nu}[\phi\star\phi\star\phi\star\phi]
\\
&&{}
-
\frac{1}{d-4}\frac{\lambda}{3!\,(4\pi)^2}
 \Bigl(
     \frac{1}{6}\eta_{\mu\nu}\partial^2[\phi\star\phi]
    -\frac{2}{3}\partial_{\mu\nu}^2[\phi\star\phi]
    +m^2\eta_{\mu\nu}[\phi\star\phi]
 \Bigr)
\\
\nonumber
\phi_0\star\partial^2_{\mu\nu}\phi_0
&=&
[\phi\star\partial^2_{\mu\nu}\phi]
-
\frac{\mu^{4-d}}{d-4}\frac{\lambda^2}{2\,3!\,(4\pi)^2}
\eta_{\mu\nu}[\phi\star\phi\star\phi\star\phi]
\\
&&{}
-
\frac{1}{d-4}\frac{\lambda}{3!\,(4\pi)^2}
 \Bigl(
     \frac{1}{6}\eta_{\mu\nu}\partial^2[\phi\star\phi]
    -\frac{2}{3}\partial_{\mu\nu}^2[\phi\star\phi]
    +m^2\eta_{\mu\nu}[\phi\star\phi]
 \Bigr)
\end{eqnarray}
we get
\begin{eqnarray}
T'_{0\mu\nu}
&=&
[T'_{\mu\nu}]
\\
\nonumber
&&{}
+\frac{\mu^{4-d}}{d-4}\frac{\lambda^2}{2\,(3!)^2(4\pi)^2}
    \eta_{\mu\nu}[\phi\star\phi\star\phi\star\phi]
    \left\{A+F-D-E\right\}
\\
\nonumber
&&{}
+\frac{1}{d-4}\frac{\lambda}{3!\,(4\pi)^2}
    \eta_{\mu\nu}\frac{m^2}{2}[\phi\star\phi]
    \left\{2A+3F-2(D+E)-2G-1\right\}
\\
\nonumber
&&{}
+\frac{1}{d-4}\frac{\lambda}{3!\,(4\pi)^2}
    \eta_{\mu\nu}\partial^2[\phi\star\phi]
    \left\{\frac{1}{6}A+2H-\frac{1}{6}(D+E)-\frac{1}{3}\right\}
\\
\nonumber
&&{}
+\frac{1}{d-4}\frac{\lambda}{3!\,(4\pi)^2}
    \partial^2_{\mu\nu}[\phi\star\phi]
    \left\{\frac{1}{3}A+\frac{2}{3}(D+E)+\frac{1}{3}\right\}.
\end{eqnarray}
Here $[T'_{\mu\nu}]$ is a finite quantity
\begin{eqnarray}
[T'_{\mu\nu}]
&=&
[T_{\mu\nu}]
+
\frac{A}{2}\left(
   [\partial_\mu\phi\star\partial_\nu\phi]
  +[\partial_\mu\phi\star\partial_\nu\phi]
\right)
+B\eta_{\mu\nu}[L^{nc}\star\phi]
+C\eta_{\mu\nu}[\phi\star{}L^{nc}]
\\
\nonumber
&&\qquad{}
+D[(\partial^2_{\mu\nu}\phi)\star\phi]
+E[\phi\star\partial^2_{\mu\nu}\phi]
+F\eta_{\mu\nu}\frac{\mu^{4-d}\lambda}{4!}
  [\phi\star\phi\star\phi\star\phi]
\\
\nonumber
&&\qquad{}
+G\eta_{\mu\nu}\frac{m^2}{2}[\phi\star\phi]
+H\eta_{\mu\nu}\partial^2[\phi\star\phi].
\end{eqnarray}
For $T'_{0\mu\nu}$ to be finite, we set all coefficients
in front of independent divergent quantities to zero and find
that
\begin{eqnarray}
E=-D-\frac{A+1}{2}
&\qquad&
F=-\frac{3A+1}{2}
\\
\nonumber
G=-\frac{3+3A}{4}
&\qquad&
H=-\frac{A-1}{8},
\end{eqnarray}
with $A$, $B$, $C$ and $D$ being independent.
Now it is easy to find
the general expression for the finite energy-momentum tensor
(i.e. the general improved energy-momentum
tensor), in
the one-loop approximation
\begin{eqnarray}
T_{0\mu\nu}^{IG}
&=&
T_{0\mu\nu}
-
\frac{1}{2}\,\eta_{\mu\nu}\,
\frac{m_0^2}{2}
\phi_0\star\phi_0
-
\frac{1}{6}\,
(\partial_\mu\partial_\nu-\eta_{\mu\nu}\partial^2)
\phi_0\star\phi_0
\\
\nonumber
&&\qquad{}
+A'\Bigl(S_{0\mu\nu}
     -\frac{1}{4}\eta_{\mu\nu}\eta^{\alpha\beta}S_{0\alpha\beta}
  \Bigr)
\\
\nonumber
&&\qquad{}
+B'\,\eta_{\mu\nu}\,L^{nc}_0\star\phi_0
+C'\,\eta_{\mu\nu}\,\phi_0\star{}L^{nc}_0
\\
\nonumber
&&\qquad{}
+D'\Bigl(
       (\partial^2_{\mu\nu}\phi_0)\star\phi_0
      -\phi_0\star\partial^2_{\mu\nu}\phi_0\Bigr)
\end{eqnarray}
Here $A'=(3A+1)/2$, $B'=B-(3A+1)/16$, $C'=C-(3A+1)/16$,
$D'=D+(A+1)/4$ are new arbitrary numbers,
$L_0^{nc}$ is given by
eq. (\ref{L0}) and $S_{0\mu\nu}$
denotes the following operators:
\begin{eqnarray}
S_{0\mu\nu}
&=&
 \frac{1}{2}\partial_\mu\phi_0\star\partial_\nu\phi_0
+\frac{1}{2}\partial_\nu\phi_0\star\partial_\mu\phi_0
-\frac{1}{6}\partial^2_{\mu\nu}(\phi_0\star\phi_0).
\end{eqnarray}
The operators standing after the arbitrary numbers $A'$, $B'$,
$C'$, $D'$ are finite, in the one-loop approximation, and do not
affect the finiteness of the energy-momentum tensor.
For simplicity we put $A'=B'=C'=D'=0$ and
define the "improved" energy-momentum tensor in the
noncommutative scalar field theory as\footnote{If $m=0$ our "improved"
energy-momentum tensor coincides with the one proposed in
\cite{0012112} on purely classical grounds,
for the energy-momentum tensor in the massless
theory to be traceless.}
\begin{eqnarray}\label{iemt}
T_{0\mu\nu}^I
&=&
T_{0\mu\nu}
-
\frac{1}{2}\,\eta_{\mu\nu}\,
\frac{m_0^2}{2}
\phi_0\star\phi_0
-
\frac{1}{6}\,
(\partial_\mu\partial_\nu-\eta_{\mu\nu}\partial^2)
\phi_0\star\phi_0.
\end{eqnarray}
It is worth pointing out that the last
"improving" term in (\ref{iemt}) coincides with the analogous term
in the corresponding commutative case \cite{Brown},
although the divergences
which they cancel differ, due to the UV/IR mixing.

\section{Summary}\label{Summary}
We have considered the renormalization of scalar
field dimension two and four composite operators and the
energy-momentum tensor, in the one-loop approximation. Using
a general theory of the renormalization of composite
operators, which is valid for both commutative and noncommutative
theories, we calculated the mixing matrix connecting
bare and renormalized composite operators and discussed its features
in noncommutative theory.
Due to noncommutativity, the number of basis dimension four operators in the case under consideration is equal 5 unlike the commutative theory where this number is 4.
The structure of the mixing matrix in the noncommutative theory 
(\ref{NCZ}) is more complicated than in the commutative one (\ref{Z}) and has no direct relation to renormalization group functions.
The mixing matrix for the composite operators at zero momentum
transfer is also found and it is shown that its matrix elements are
connected with the renormalization group functions of the
noncommutative theory.

We have studied the problem of the
renormalization of the canonically defined energy-momentum tensor.
It turns out that the latter is not finite, hence we are forced to consider its improvement.
We defined the improved energy-momentum tensor as a sum of the canonical one
and an arbitrary linear combination of suitable composite operators.
Requiring finiteness leads us to determine
conditions on the coefficients of the above
linear combination and allows us to fix it. As a result, we have derived the
``improved'' energy-momentum tensor in noncommutative scalar field theory.

\section*{Acknowledgements}
This work was supported in part
by the European Community's Human Potential
Programme under the contract HPRN-CT-2000-00131 Quantum Spacetime,
the INTAS-00-0254 grant, the NATO Collaborative Linkage Grant PST.CLG.979389,
the RFBR grant, project No.\ 03-02-16193,
the joint RFBR-DFG grant, project No.\ 02-02-04002, the
DFG grant, project No.\ 436 RUS 113/669
and the grant PD02-1{.}2-94 of Russian Ministry of Education.
I.L.B. is very grateful to Center of String
and Particle Theory at the University of Maryland where
this work was undertaken, for partial support and S.J. Gates for
hospitality. He is also grateful for partial support to INFN,
Laboratori Nazionali di Frascati where the work was completed.

\appendix

\section*{Appendix}
\section*{Improved energy-momentum tensor and global conservation
law}
\renewcommand{\theequation}{A.\arabic{equation}}
\setcounter{equation}{0}

In this appendix we examine the conservation of energy-momentum in the
theory with the "improved" energy-momentum tensor (\ref{iemt}).
The general situation in noncommutative field theories is as follows.
In deriving conserved quantities with the help of Noether's theorem,
the cyclicity property of the star product (\ref{cp}), which is proved
by integration by parts, is used.
Therefore, in the equation which arise in Noether's theorem
\begin{equation}
\label{A1}
\epsilon^k\int\partial_\mu J^\mu_k\,d^4x=0,
\end{equation}
with $\epsilon^k=const$ being arbitrary global parameters of the
symmetry transformation and $J^\mu_k$ being currents, the integration
region in noncommutative theories is the whole space-time, unlike
commutative theories where the integration region is still arbitrary,
because integration by parts is not used.
As a result, in contrast with the commutative case, we do not have
local conservation laws in noncommutative theory.
However, the quantity $\partial_\mu{}J^\mu$ must vanish at
$\theta\to0$, since there exists a smooth
commutative limit at the classical level.
Therefore, the most we may have is
(see e.g. \cite{0008057})
\begin{eqnarray}
&&\partial_\mu J^\mu
=
\Bigl\{f,g\Bigr\},
\\&&\qquad
\Bigl\{f,g\Bigr\}\equiv{}f\star{}g-g\star{}f,
\end{eqnarray}
with $f$ and $g$ being some functions of the fields. In this case, we have
\begin{eqnarray}
\label{div}
\int\partial_\mu{}J^\mu\,d^{3}x
&=&
\int\Bigl\{f,g\Bigr\}d^{3}x.
\end{eqnarray}
The r.h.s. of (\ref{div}) vanishes only in the case
$\theta^{0i}=0$, and then the charge $Q=\int{}J^0\,d^{3}x$ is
conserved. Thus, in noncommutative field theories
there are only global conservation laws\footnote{An attempt to
construct the locally conserved energy-momentum tensor in
noncommutative scalar field theory was made in the work
\cite{0104244}.} and only when $\theta^{0i}=0$.

This situation may also be explained as follows.
In calculating (\ref{A1}) in the case of only spatial
noncommutativity $\theta^{0i}=0$, integrating by parts is
carried out only in spatial directions and not in the
time coordinate.
So, the time integration region is still arbitrary, in this case,
and we can write
\begin{eqnarray}
\int\partial_\mu{}J^{\mu}d^3x
&=&
\partial_0\int{}J^0d^3x=0.
\end{eqnarray}
Thus, we have the conserved quantity $Q=\int{}J^0d^3x$.

Let us turn to noncommutative scalar field theory. If we choose the
energy-momentum tensor as in (\ref{bemt}), it satisfies the relation
\begin{eqnarray}
\partial^\mu T_{\mu\nu}
=
\frac{\lambda}{4!}
\Bigl\{\bigl\{\phi,\partial_\nu\phi\bigr\},\phi\star{}\phi\Bigr\}.
\end{eqnarray}
This means that a global conservation law exists.
However, such choice of $T_{\mu\nu}$ is not finite.
If we improve the energy-momentum tensor and make
it finite (\ref{iemt}), then
\begin{eqnarray}
\label{diviemt}
\partial^\mu T_{\mu\nu}^I
&=&
\frac{\lambda}{4!}
\Bigl\{\bigl\{\phi,\partial_\nu\phi\bigr\},\phi\star{}\phi\Bigr\}
-
\frac{1}{4}\, m^2 \partial_\nu (\phi\star\phi).
\end{eqnarray}
From (\ref{diviemt}) we see that, after integrating over the
whole space, the last term in (\ref{diviemt}) disappears if $\nu$
is a spatial index and the field $\phi$ has the proper
asymptotic behaviour. So, the momentum is always conserved, but
the energy is conserved only for a vanishing mass. Hence, in the
theory under consideration we may construct an energy-momentum
tensor which is finite and leads to conserved energy and
momentum only in a massless theory.

However, there exists another way to define global
conservation laws.
Let us introduce the quantity
\begin{eqnarray}
P_{\nu}=\int{}T^0_{\nu}d^{3}x
\label{cemv}
\end{eqnarray}
with $T^\mu_{\nu}$ being the canonical energy-momentum tensor
(\ref{bemt}).
It is evident that $P_\nu$ in (\ref{cemv}) is conserved in time
$\partial^0P_{\nu}=0$ for the spatial noncommutativity.
Next, we show that in this case the quantity (\ref{cemv}) is a finite
operator, at least in the one-loop approximation.

The operator $T_{\mu\nu}$ (\ref{bemt}) is a linear combination
of five composite operators.
As a result, the operator $P_{\nu}$ in (\ref{cemv}) is a linear
combination of these composite operators at zero momentum
transfer.
We study the renormalization of the operator $P_{\nu}$ in (\ref{cemv}).
Let us begin with the operator
\begin{math}
\frac{m^2_0}{2}\int\phi_0\star\phi_0\,d^{d-1}x.
\end{math}
In order to renormalize such an operator, we put $J_1(x)=\delta(x^0-t)$ in
the integrand of (\ref{1op}).
Then, this corresponds to $J_1\sim\delta(\vec{k})$ in the expression
(\ref{f1b}), which leads to $p_\mu\theta^{\mu\nu}k_\nu=0$ in
(\ref{mf2}), and we find that the number of planar diagrams is
twice more than in the case of the renormalization of the
operator at arbitrary momentum transfer and so the counterterm
is also two times larger than (\ref{f2ct}).
Thus, we have
\begin{eqnarray}
\frac{m_0^2}{2}\int\phi_0\star\phi_0\,d^{d-1}x
&=&
\frac{m^2}{2}\left[\int\phi\star{}\phi\,d^{d-1}x\right].
\label{f^2}
\end{eqnarray}

Analogous considerations may be carried out for the other
operators and we get
\begin{eqnarray}
\nonumber
\frac{\lambda_0}{4!}
\int\phi_0\star\phi_0\star\phi_0\star\phi_0\,d^{d-1}x
&=&
\left(1+\frac{1}{d-4}\frac{2/3\lambda}{(4\pi)^2}\right)
  \frac{\mu^{4-d}\lambda}{4!}
\left[
  \int\phi\star\phi\star\phi\star\phi\,d^{d-1}x\right]
\\
&&{}
+
\frac{1}{d-4}\frac{2/3\lambda}{(4\pi)^2}
\frac{m^2}{2}\left[\int\phi\star\phi\,d^{d-1}x\right],
\label{f^4}
\\
\nonumber
\int\partial_\mu\phi_0\star\partial_\nu\phi_0\,d^{d-1}x
&=&
\left[\int\partial_\mu\phi\star\partial_\nu\phi\,d^{d-1}x\right]
\\&&{}
\label{pf^2}
\hspace*{-20mm}
+
\frac{\mu^{4-d}}{d-4}
\frac{\lambda^2}{(3!)^2\,(4\pi)^2}\,
\eta_{\mu\nu}
\left[\int\phi\star{}\phi\star{}\phi\star{}\phi\,d^{d-1}x\right]
+
\\&&{}
\nonumber
\hspace*{-20mm}
+
\frac{1}{d-4}\frac{\lambda/3}{(4\pi)^2}
\Bigl(
\frac{1}{6}\eta_{\mu\nu}\partial^2_{00}
+
\frac{1}{3}\delta^0_\mu\delta^0_\nu\partial^2_{00}
+
m^2\eta_{\mu\nu}
\Bigr)
\left[\int\phi\star\phi\,d^{d-1}x\right].
\end{eqnarray}
It should be noted here that in the case of spatial
noncommutativity $\theta^{0i}=0$ we have got the same
renormalization constants as in the case of the
renormalization of
the operators at zero momentum transfer (\ref{mff}, \ref{ffff},
\ref{fppf}).
This happened, owing to the fact that such a noncommutativity does not
affect the time components of variables, and thus there exists a
smooth limit $p_0\to0$ ($p_0$ being the time component of the
transferred momentum) at the quantum level.
So, it makes no difference, whether the operator is integrated over
time or not.

Let us turn now to the energy-momentum vector $P_\nu$ in (\ref{cemv}).
Substituting eqs. (\ref{f^2}, \ref{f^4}, \ref{pf^2}) into the r.h.s.
of (\ref{cemv}) one gets
\begin{eqnarray}
P_{\nu}
&=&
\frac{1}{2}
\left[\int\partial_0\phi\star\partial_\nu\phi\,d^{3}x\right]
+
\frac{1}{2}
\left[\int\partial_\nu\phi\star\partial_0\phi\,d^{3}x\right]
-
\frac{1}{2}\,\eta_{0\nu}
\left[\int\partial_\alpha\phi\star\partial^\alpha\phi\,d^{3}x\right]
\nonumber
\\
&&{}
+
\eta_{0\nu}\,
\frac{m^2}{2}\left[\int\phi\star{}\phi\,d^{3}x\right]
+
\eta_{0\nu}\,
\frac{\lambda}{4!}
\left[
  \int\phi\star\phi\star\phi\star\phi\,d^{3}x\right].
\label{remv}
\end{eqnarray}
We see that (\ref{remv}) is expressed in terms of renormalized
composite operators and hence is also finite.
Thus, the quantity $P_{\nu}$ in (\ref{cemv}) may be considered as
the energy-momentum vector of the field $\phi$ in the
noncommutative theory.

\end{document}